\documentclass[twocolumn,showpacs,prl]{revtex4}
\usepackage{graphicx}
\begin{document}
\title{Quantum phase transitions in a two-dimensional quantum XYX model: \\Ground-state fidelity and entanglement}
\author{Bo Li, Sheng-Hao Li, and Huan-Qiang Zhou}
\affiliation{Centre for Modern Physics and Department of Physics,
Chongqing University, Chongqing 400044, The People's Republic of
China}
\date{\today}

\begin{abstract}
A systematic analysis is performed for quantum phase transitions in
a two-dimensional anisotropic spin $1/2$ anti-ferromagnetic XYX
model in an external magnetic field. With the help of an innovative
tensor network algorithm, we compute the fidelity per lattice site
to demonstrate that the field-induced quantum phase transition is
unambiguously characterized by a pinch point on the fidelity
surface, marking a continuous phase transition. We also compute an
entanglement estimator, defined as a ratio between the one-tangle
and the sum of squared concurrences,  to identify both the
factorizing field and the critical point, resulting in a
quantitative agreement with quantum Monte Carlo simulation. In
addition, the local order parameter is ``derived'' from the tensor
network representation of the system's ground state wave functions.
\end{abstract}
 \maketitle

Quantum critical phenomena are crucial in our understanding of the
underlying physics in quantum many-body systems, especially in
condensed matter systems,  due to their relevance to high-$T_{c}$
superconductors, fractional quantum Hall liquids, and quantum
magnets~\cite{sachdev,wen}. The latest advances in this area arise
from quantum information science. Indeed, various entanglement
measures have been widely applied to study condensed matter systems.
Remarkably, for one-dimensional (1D) quantum systems, von-Neumann
entropy, as a bipartite entanglement measure, turns out to be a good
criterion to judge whether or not a system is at
criticality~\cite{preskill,
osborne,vidal_ent,korepin,levin,entanglement,ent_review}. On the
other hand, fidelity, another basic notion in quantum information
science, has demonstrated to be fundamental in characterizing phase
transitions in quantum many-body
systems~\cite{zanardi,zhou,zhou_order,fidelity}. This adds a new
routine to explore quantum criticality in condensed matter physics
from a quantum information perspective.

However, with only a few notable exceptions~\cite{iPEPS-fidelity,
compass}, not much work has been done for two-dimensional (2D)
quantum systems, due to great computational challenges. In fact,
despite the existence of well-established numerical algorithms, such
as exact diagonalization, quantum Monte Carlo (QMC), the density
matrix renormalization group (DMRG) and series expansions, drawbacks
become obvious when one deals with frustrated spin systems. A
typical example is the QMC, which suffers from the notorious sign
problem. However, a promising progress, inspired by new concepts
from quantum information science, has been made in classical
simulations of quantum many-body systems. The algorithms are based
on an efficient representation of the system's wave functions
through a tensor network. In particular, matrix product states
(MPS)~\cite{MPS,TI_MPS,PBC_MPS}, a tensor network already present in
DMRG, are used in the time-evolving block decimation (TEBD)
algorithm to simulate time evolution in 1D quantum lattice
systems~\cite{TEBD,iTEBD}, whereas projected entangled-pair states
(PEPS) constitute the basis to simulate 2D quantum lattice
systems~\cite{PEPS,iPEPS}.

The aim of this paper is to show that the fidelity per lattice site,
first introduced in Ref.~\cite{zhou}, is able to unveil quantum
criticality for a 2D anisotropic spin $1/2$ anti-ferromagnetic XYX
model in an external magnetic field. This is achieved by exploiting
tensor network algorithms, i.e., innovative algorithms inspired by
the latest achievements in our understanding of quantum
entanglement~\cite{PEPS,iPEPS}. We show that the field-induced
quantum phase transition is unambiguously characterized by a pinch
point on the fidelity surface, marking a continuous phase
transition.  In addition, we compute an entanglement estimator,
defined as a ratio between the one-tangle and the sum of squared
concurrences (for all the pairwise entanglement between two spins),
to identify both the factorizing field and the critical point,
resulting in consistent conclusions as drawn from the fidelity
approach, with an extra result about a factorizing field $h_{f}$.
Our results are compared to those of the QMC simulation by Roscilde
\textit{et al.}~\cite{XYX-QMC} for the model. We stress that the QMC
simulation is carried out for a system on a finite square lattice at
very low but finite temperatures, whereas our simulation is directly
performed for an infinite system at zero temperature.

{\it Quantum XYX model.} We consider the 2D antiferrmagnetic
spin-1/2 XYX model in a uniform z axis external magnetic field:
\begin{equation}
H=J \sum_{< i,j >} (
S^{x}_{i}S^{x}_{j}+\Delta_{y}S^{y}_{i}S^{y}_{j}+S^{z}_{i}S^{z}_{j})
+\sum_{i}hS^{z}_{i}, \label{Hamt}
\end{equation}
where $J>0$ is the exchange coupling, $<i,j >$ runs over all the
possible pairs of the nearest neighbors on a square lattice, and $h$
is the external magnetic field. From the non-commutativity of the
spin-1/2 Pauli operators, XYX model is expected to undergo a
continuous quantum phase transition, with the same universality
class as the 2D quantum Ising model in a transverse field. $
\Delta_{y}<1$ and $ \Delta_{y}>1$ correspond to easy-plane (EP) and
easy-axis (EA) behaviors, respectively. The ordered phase in the EP
(EA) case arises from spontaneous symmetry breaking along the x (y)
direction, with a finite value of the order parameter, i.e., the
magnetization $m_{x}$ ($m_{y}$) below the critical field $h_{c}$. In
Ref.~\cite{XYX-QMC}, the QMC simulation was exploited to discuss the
connection between quantum phase transitions and entanglement
measures, where an entanglement estimator, defined as the ratio
between the one-tangle and the sum of squared concurrences, was
systematically analyzed to signal a quantum critical point.

{\it The infinite projected entangled-pair state (iPEPS) method.}
Let us briefly recall the iPEPS algorithm. Consider a finite
two-dimensional square lattice where each site, labeled by a vector
$\vec{r} = (x,y)$, is represented by a local Hilbert space
$V^{[\vec{r}]} \cong {C}^d$ of finite dimension $d$. Let a vector
$|\Psi\rangle$ denote a pure state in the (global) Hilbert space and
the operator $H = \sum_{\vec{r},\vec{r}'} h^{[\vec{r}\vec{r}']}$ be
a Hamiltonian with the nearest neighbor interactions on the lattice.
Each lattice site has been represented by a tensor $A^{[\vec{r}]}$,
so a PEPS for the state $|\Psi\rangle$ consists of a set of tensors
$A^{[\vec{r}]}$. The tensor $A^{[\vec{r}]}_{sudlr}$ is made of
complex numbers labeled by one {\em physical} index $s$ and four
\textit{inner} indices $u$, $d$, $l$ and $r$. The physical index
runs over a basis of $V^{[\vec{r}]}$, so that $s=1,\cdots,d$,
whereas each inner index takes $D$ values, where $D$ is some bond
dimension, and connects the tensor with the tensors in the nearest
neighbor sites. Thus, in a lattice with $N$ sites, a PEPS depends on
$O(ND^4d)$ parameters~\cite{PEPS}.

Now we move to a system defined on an infinite square lattice and
assume that both $|\Psi\rangle$ and $H$ are invariant under shifts
by two lattice sites. We exploit this invariance to store the iPEPS
using only two different tensors $A$ and $B$. Given an iPEPS for a
state $\Psi_0$ (e.g., a product state), the iPEPS algorithm allows
to perform an evolution in imaginary time to compute a ground state
wave function of a given Hamiltonian $H$, $ |\Psi_{\tau}\rangle =
{e^{-H\tau} |\Psi_0\rangle}/||e^{-H\tau}
|\Psi_0\rangle||$~\cite{iPEPS}.

A contraction process, which is related to an evolution task, is
done in order to get the effective environment for a pair of tensors
$A$ and $B$~\cite{iPEPS}. In practice, a global optimization problem has been
reduced to a local two site optimization problem. Two new tensors
$A'$ and $B'$ could be computed by a sweep technique~\cite{TI_MPS},
originally devised for an MPS algorithm applied to 1D quantum
systems with periodic boundary conditions~\cite{PBC_MPS}.

\begin{figure}
  \begin{center}
    \includegraphics[width=8.5cm]{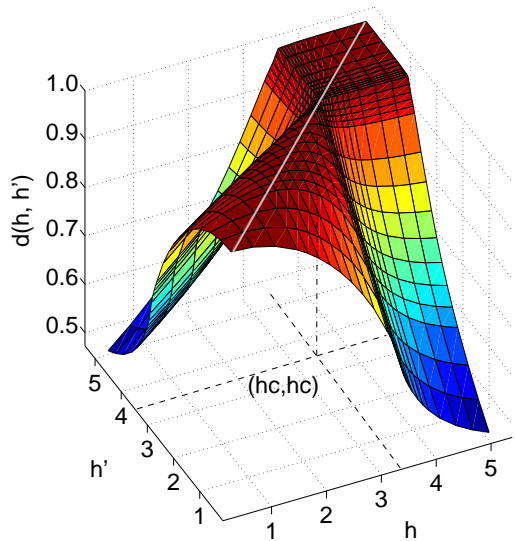}
  \end{center}
   \vspace*{-1.0cm}
  \caption{(color online) The  fidelity per lattice site $d(h,h')$, as a
  function of $h$ and $h'$ for two ground states of the two-dimensional 
  quantum XYX model. This defines a 2D fidelity surface embedded in a
three-dimensional Euclidean space. A continuous phase transition
point $h_{c} \approx 3.489$ is characterized as a pinch point
($h_{c}$,$h_{c}$) on the fidelity surface, as argued in
Ref.~\cite{zhou}.  Here we have taken the bond dimension $D=2$.
The gray line denotes the normalization: $d(h,h) = 1$.}
  \label{fidelity}
\end{figure}

{\it The fidelity per lattice site.} Consider a finite 2D square
lattice system described by Eq.~(\ref{Hamt}), with the external
magnetic field $h$ as a control parameter. For two different ground
states of the system, $\Psi(h)$ and $\Psi(h')$, corresponding to two
different values $h$ and $h'$ of the control parameter,
respectively, the fidelity per lattice site
 $d$ is defined as:
\begin{equation}
\ln d(h,h') =   \frac {\ln F(h,h')}{N}, \label{d}
\end{equation}
where $N$ is the system size, and $F(h,h') \equiv | \langle
\psi(h')|\psi(h)\rangle|$ is the ground state fidelity. The fidelity
per lattice site $d$ depicts how fast the fidelity goes to zero when
$N$ gets large.  Remarkably, the fidelity per lattice site $d$ is
well defined in the thermodynamic limit:
\begin{equation}
\ln d(h,h') = \lim_{N \rightarrow \infty} \frac {\ln F(h,h')}{N}.
\label{dinf}
\end{equation}
It satisfies the properties inherited from  fidelity $F(h,h')$: (i)
normalization $d(h,h)=1$; (ii) symmetry $d(h,h')=d(h',h)$; and (iii)
range $0 \le d(h,h')\le 1$.

As shown in Ref.~\cite{zhou}, the fidelity per lattice site
$d(h,h')$ succeeds in capturing nontrivial information about stable
and unstable fixed points along renormalization group flows.
Specifically,  the fidelity surface, defined by the fidelity per
lattice site $d(h,h')$ as a 2D surface embedded in a
three-dimensional Euclidean space, exhibits singularities when
$h=h_c$ or $h'=h_c$. That is, $d(h,h')$
exhibits singular behaviors when $h$ crosses $h_c$ for a fixed $h'$,
or $h'$ crosses $h_c$ for a fixed $h$. Therefore, a phase transition
point $h_c$ is characterized as a \textit{pinch
point}~\cite{pinch_point} $(h_c,h_c)$ for \textit{continuous} QPTs,
i.e., the intersection of two singular lines $h=h_c$ and $h'=h_c$.

The fidelity per lattice site may be computed from the iPEPS
representation of the ground state wave functions, following the
transfer matrix approach described in Ref.~\cite{iPEPS-fidelity}. We
plot $d(h,h')$ in Fig.~\ref{fidelity}, computed with the help of the
iPEPS algorithm~\cite{iPEPS} with bond dimension $D= 2$ (the result
for $D=3$ is very similar to that for $D= 2$). A pinch point
on the fidelity surface defined by $d(h,h')$ as a function of $h$
and $h'$ clearly indicates a second order phase transition. In
addition, the two stable fixed points at $h=0$ and $h=\infty$ are
characterized as the global minima of the fidelity surface (for a
fixed $\Delta_y$).

\begin{figure}[ht]
  \begin{center}
    \includegraphics[width=8.5cm]{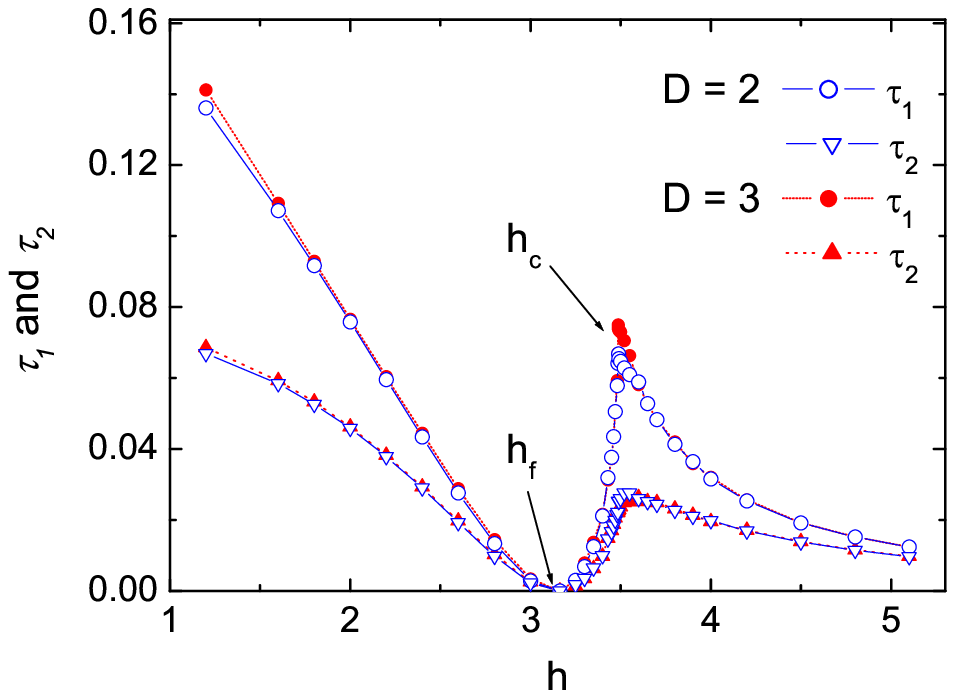}
  \end{center}
   \vspace*{-1.5cm}
  \begin{center}
  \includegraphics[width=8.5cm]{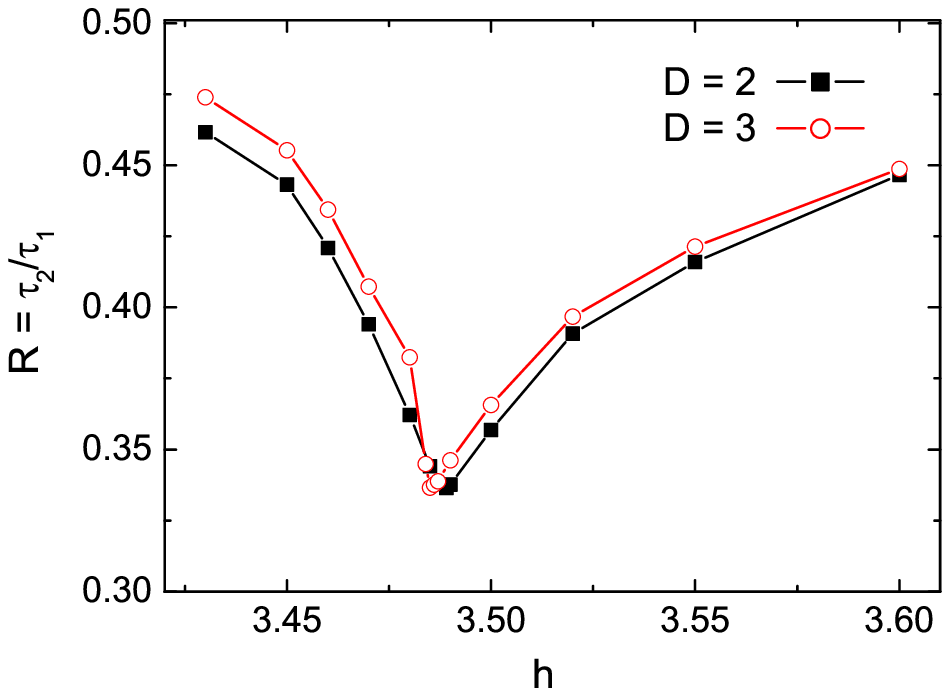}
  \end{center}
 \vspace*{-1.0cm}
  \caption{(color online) Upper panel: The one-tangle $\tau_{1}$
  and the sum of squared concurrences
 $\tau_2$ as a function of an applied external field $h$ for the 2D
 quantum XYX  model with $\Delta_{y}=0.25$. The data are presented for both $D= 2$ and $D= 3$.
 The factorizing field, at which the one-tangle $\tau_1$ vanishes, is indicated by an arrow
 labeled by $h_{f}$ around 3.162. Above the factorizing field $h_{f}$, a steep
 increase of $\tau_1$ and $\tau_2$ reflects a rapid increase of entanglement
 around a critical point $h_{c} \approx 3.489 \;(D=2)$ and $h_{c} \approx 3.485 \;(D=3)$.
 Lower panel: An entanglement ratio $R=\tau_2 / \tau_1$, exhibits a cusp, a
signal of a continuous phase transition,  at a critical point $h_{c}$ obtained
using the infinite projected entangled-pair state (iPEPS) algorithm with $D= 2$ and $3$.  
This is in an agreement with that resulted from the fidelity approach.
}
  \label{one-tangle}
\end{figure}

{\it The one-tangle and the concurrence.} We now exploit the iPEPS
algorithm to extract the ground state entanglement properties of the
quantum XYX model on an infinite square lattice. We first compute
the one-tangle $\tau_{1}$,  defined as $\tau_{1}=4 \det \rho^{(1)}$,
where $\rho^{(1)}$ is the single site reduced density matrix, for a
specific value $\Delta_{y}=0.25$ with $D=2$ and $3$. However, we stress that, our
discussion, although only confined to the $\Delta_{y}=0.25$ case, is
actually quite generic and applies to all other values of
$\Delta_{y}<1$. The one-tangle reflects the entanglement between a
single site and the rest of the system. Note that there exists a factorizing
field, at which the one-tangle
$\tau_{1}$ vanishes~\cite{vot}. The exact theoretical value of the factorizing
field  $h_{f}=2\sqrt{2(1+\Delta_{y})}$ for  $\Delta_{y}=0.25$
 is approximately 3.162, consistent with the iPEPS
 results up to four digits. When the external field is increased
 beyond the factorizing field $h_{f}$,
 a cusp occurs for an entanglement ratio $R=\tau_2 / \tau_1$ at a critical
 point $h_{c}$, where  $\tau_2$ denotes the sum of squared
concurrences (for all the pairwise entanglement between two spins).
The cusp in the entanglement ratio can be regarded as a
 signal of a quantum phase transition.
 In Fig.~\ref{one-tangle},  both the one-tangle $\tau_{1}$ and
 the sum of squared concurrences $\tau_2$, obtained from the iPEPS
 with $D=3$, are greater than those for $D=2$.  This is due to the fact that for larger $D$,
 the PEPS representation  accommodates more entanglement.
 The critical point $h_{c}\approx 3.485$
from the iPEPS for $D=3$ is smaller than that for
 $D=2$  ($h_{c}\approx 3.489$). The shift is quite small, so one may expect that  the critical point for  small $D$
 does not deviate significantly from  the converged result with large $D$.
Our result indicates that the iPEPS algorithm is able to
capture the entanglement properties, and gives rise to the
 results consistent with the QMC simulation.

{\it Local order parameter.} The efficient tensor network
representation of the system's ground state wave functions makes it
possible to extract an (optimized) local order parameter, according
to a general scheme advocated in Ref.~\cite{zhou_order}. In fact,
once the critical field $h_c$ is determined, one may choose two
representative ground states, one for an external magnetic field $h$
less than the critical field $h_c$ and the other for an external
magnetic field $h$ greater than the critical field $h_c$. Then the
reduced density matrix $\rho^{(1)}$ for a single lattice site in an
infinite-size lattice is computed for two different values of the
external magnetic field $h$, corresponding to $h > h_c$ and $h <
h_c$, respectively. It is readily found that the one-site reduced
density matrix $\rho^{(1)}$ displays different nonzero-entries
structures in two phases, with $\langle S_x \rangle$ being zero for
$h>h_c$ and nonzero for $h<h_c$. Also note that the $Z_2$ symmetry
is spontaneously broken, since the reduced density matrix
$\rho^{(1)}$ does not commute with the symmetry generating operator
in the symmetry-broken phase $h<h_c$. This implies the existence of
a local order parameter: $m_x=\langle \psi(h)|S_{x}|\psi(h)\rangle$,
characterizing the second order phase transition that belongs to the
same universality class as that of the 2D quantum Ising model in a
transverse field.

Another interesting local observable $m_z=\langle
\psi(h)|S_{z}|\psi(h)\rangle$ is also carefully investigated, which
exhibits singularities at the critical field $h_c$. If the external
transverse magnetic field $h$ is raised from below the critical
field $h_{c}$, then the $z$ magnetization is a monotonic curve and
gradually reaches a saturated value.

\begin{figure}
  \begin{center}
    \includegraphics[width=8.5cm]{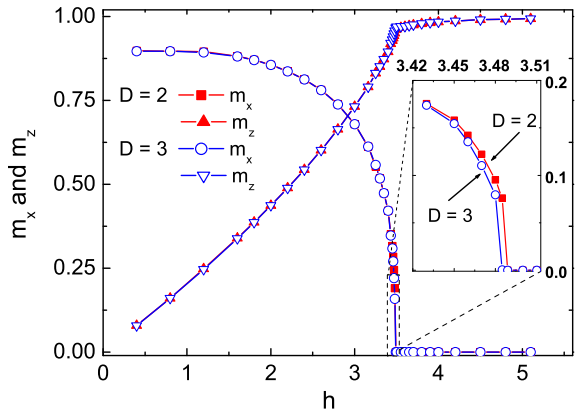}
  \end{center}
   \vspace*{-1.0cm}
  \caption{(color online) The $x$ magnetization $m_x(h)$ in the ground state
  $|\Psi(h)\rangle$ of the 2D quantum XYX model ($\Delta_{y}=0.25$),
  a local order parameter, is readily read off from the infinite projected
  entangled-pair state (iPEPS) representation of the ground state wave functions
  with $D=2$ and $3$.  Dashed lines are a guide to the eye.
  For the external magnetic field $h$ less than the critical magnetic
  field $h_c$, the order parameter takes a  non-zero value and
  decays as $h$ increases. Once the field is greater than the critical
  field $h_c$, the order parameter becomes zero.  Notice that,
  around $h_{c}$, the derivative of the z magnetization $m_{z}(h)$
  changes rapidly. Thus it exhibits singularities at the critical field
 $h_c$.}
  \label{orderparameter}
\end{figure}

We plot $m_x$  and $m_z$ as a function of
the external magnetic field $h$ for the 2D quantum XYX model with
$\Delta_{y}=0.25$ in Fig.~\ref{orderparameter}.  We have presented
the data for both  $D=2$ and $3$, with the truncation dimension $\chi$
 for iMPS used in the contraction of the iPEPS representation up to 30.
On the other hand, although the factorizing field can be readily located in
Fig.~\ref{one-tangle}, there are no unusual features appearing in
$m_x$ and $m_z$ around this point.

{\it Conclusions.} We have performed a systematic
analysis of both the ground state fidelity and entanglement for the
2D quantum XYX model. This is achieved by computing the ground state
wave functions by means of the iPEPS algorithm. The results are
compared with those obtained from the QMC
simulation~\cite{XYX-QMC}, giving rise to a remarkable quantitative
agreement for the factorizing field. From a fidelity perspective, a
connection between a pinch point on the fidelity surface and a
continuous phase transition point has been demonstrated,  thus
allowing us to determine the ground state phase diagram of the 2D
quantum XYX model. Once this has been done, one may read off the
local order parameter from representative states in terms of the
iPEPS representation of the ground state wave functions.

An interesting point worth to be mentioned is that, in contrast to
entanglement measures, the fidelity per lattice site fails to locate
a factorizing field. However, one may resort to a closely related
quantity, i.e.,  the so-called geometric entanglement introduced in
Ref.~\cite{gm}, to locate it, as long as such a factorizing field
exists for a system considered~\cite{geometric}.

This work is supported in part by the National Natural Science
Foundation of China (Grant Nos. 10774197 and 10874252).


\begin{thebibliography}{99}

\bibitem{sachdev} S. Sachdev, \textit{Quantum Phase Transitions},
Cambridge University Press, 1999, Cambridge.

\bibitem{wen} X.-G. Wen, \textit{Quantum Field Theory of Many-Body
Systems}, Oxford University Press, 2004, Oxford.

\bibitem{preskill} J. Preskill, J. Mod. Opt. \textbf{47}, 127
(2000).

\bibitem{osborne} T.J. Osborne and M.A. Nielsen, Phys. Rev. A
\textbf{66}, 032110 (2002); A. Osterloh \textit{et al.}, Nature
\textbf{416}, 608 (2002).

\bibitem{vidal_ent} G. Vidal \textit{et al.}, Phys. Rev. Lett.
\textbf{90}, 227902 (2003); G. Vidal \textit{et al.}, Phys. Rev. Lett.
\textbf{99}, 220405 (2007); G. Evenbly and G. Vidal, arXiv:0710.0692.

\bibitem{korepin} V.E. Korepin, Phys. Rev. Lett. \textbf{92}, 096402 (2004);
G.C. Levine, Phys. Rev. Lett. \textbf{93}, 266402 (2004);
G. Refael and J.E. Moore, Phys. Rev. Lett. \textbf{93}, 260602 (2004);
P. Calabrese and J. Cardy, J. Stat. Mech. P06002 (2004).

\bibitem{levin} A. Kitaev and J. Preskill, Phys. Rev. Lett.
\textbf{96}, 110404 (2006); M. Levin and X.-G. Wen,
Phys. Rev. Lett. \textbf{96}, 110405 (2006).

\bibitem{entanglement} F. Verstraete, M.A. Martin-Delgado, and J.I. Cirac,
Phys. Rev. Lett. \textbf{92}, 087201 (2004);
W. D¨ur \textit{et al.}, Phys. Rev. Lett. \textbf{94}, 097203 (2005);
H. Barnum \textit{et al.}, Phys. Rev. Lett. \textbf{92}, 107902 (2004).

\bibitem{ent_review} L. Amico \textit{et al.}, Rev. Mod. Phys.
\textbf{80}, 517 (2008).

\bibitem{zanardi} P. Zanardi and N. Paunkovi\'{c},
Phys. Rev. E \textbf{74}, 031123 (2006).

\bibitem{zhou} H.-Q. Zhou and J.P. Barjaktarevi$\check{\rm c}$,
J. Phys. A: Math. Theor. \textbf{41}, 412001 (2008); H.-Q. Zhou,
J.-H. Zhao, and B. Li, J. Phys. A: Math. Theor. \textbf{41}, 492002
(2008); H.-Q. Zhou, arXiv:0704.2945.

\bibitem{zhou_order} H.-Q. Zhou, arXiv:0803.0585.

\bibitem{fidelity} P. Zanardi, M. Cozzini, and P. Giorda,
arXiv:cond-mat/ 0606130; N. Oelkers
and J. Links, Phys. Rev. B \textbf{75}, 115119 (2007); M. Cozzini,
R. Ionicioiu, and P. Zanardi, arXiv:cond-mat/0611727; L. Campos Venuti and
P. Zanardi, Phys. Rev. Lett. \textbf{99}, 095701 (2007); P.
Buonsante and A. Vezzani, Phys. Rev. Lett. \textbf{98}, 110601
(2007); W.-L. You, Y.-W. Li, and S.-J. Gu, Phys. Rev. E \textbf{76},
022101 (2007); S.-J. Gu \textit{et al.}, arXiv:0706.2495; M.F. Yang,
arXiv:0707.4574; Y.-C. Tzeng and M.-F. Yang, arXiv:0709.1518;
S. Chen \textit{et al.}, Phys. Rev. A \textbf{77}, 032111 (2008);
L. Campos Venuti \textit{et al.}, Phys. Rev. B \textbf{78}, 115410 (2008);
J.O. Fjaerestad, J. Stat. Mech. P07011, (2008).

\bibitem{iPEPS-fidelity} H.-Q. Zhou,  R. Or\'us, and G. Vidal,
Phys. Rev. Lett. \textbf{100}, 080601 (2008).

\bibitem{compass} R. Or\'us, A.C. Doherty, and G. Vidal, arXiv:0809.4068

\bibitem{MPS} M. Fannes, B. Nachtergaele, and R.F. Werner,
Comm. Math. Phys. \textbf{144}, 443 (1992); J. Funct. Anal.
\textbf{120}, 511 (1994); S. \"Ostlund and S. Rommer, Phys. Rev.
Lett. \textbf{75}, 3537 (1995).

\bibitem{TI_MPS} D. Perez-Garcia \textit{et al.}, Quantum Inf. Comput. \textbf{7}, 401 (2007),
arXiv:quant-ph/0608197.

\bibitem{PBC_MPS}F. Verstraete, D. Porras, and J.I. Cirac,
Phys. Rev. Lett. \textbf{93}, 227205 (2004).

\bibitem{TEBD} G. Vidal, Phys. Rev. Lett. \textbf{91}, 147902 (2003);
G. Vidal, Phys. Rev. Lett. \textbf{93}, 040502 (2004).

\bibitem{iTEBD} G. Vidal, Phys. Rev. Lett. \textbf{98}, 070201 (2007).

\bibitem{PEPS} F. Verstraete and J.I. Cirac, arXiv:cond-mat/0407066;
V. Murg, F. Verstaete, and J.I. Cirac,  Phys. Rev. A \textbf{75}, 033605 (2007).

\bibitem{iPEPS} J. Jordan \textit{et al.}, Phys. Rev. Lett. \textbf{101}, 250602 (2008).

\bibitem{XYX-QMC} T. Roscilde \textit{et al.}, Phys. Rev. Lett. \textbf{94}, 147208 (2005).

\bibitem{pinch_point} The terminology ``pinch point'' was first introduced in Ref.~\cite{zhou}.

\bibitem{vot} This is due to the fact that there is no entanglement
in a factorized state.

\bibitem{gm} T.-C. Wei and P.M. Goldbart, Phys. Rev. A \textbf{68}, 042307 (2003);
T.-C. Wei \textit{et al.}, Phys. Rev. A \textbf{71}, 060305(R) (2005).

\bibitem{geometric}
The computation of the geometric entanglement per lattice site from
a tensor network representation of the system's ground state wave
functions is discussed in Q.-Q. Shi, R. Or\'us, J.O. Fjaerestad and H.-Q. Zhou,
arXiv:0901.2863; For matrix product states, see also, R. Or\'us,  Phys. Rev. Lett.
 \textbf{100}, 130502 (2008).

\end{thebibliography}
\end{document}